\documentclass[lettersize,journal]{IEEEtran}
\PassOptionsToPackage{hyphens}{url}
\usepackage{xcolor}

\usepackage{makecell}
\usepackage{longtable}
\usepackage{booktabs}
\usepackage{float}
\usepackage{amsmath,amsfonts}
\usepackage{algorithmic}
\usepackage{algorithm}
\usepackage{array}
\usepackage[caption=false,font=normalsize,labelfont=sf,textfont=sf]{subfig}
\usepackage{tabularray}
\usepackage{textcomp}
\usepackage{stfloats}
\usepackage{url}
\usepackage{verbatim}
\usepackage{graphicx}
\usepackage{cite}
\usepackage{multirow} 
\usepackage[colorlinks,linkcolor=red,anchorcolor=green,citecolor=blue]{hyperref}
\usepackage{cleveref}
\usepackage{xcolor}
\usepackage{colortbl}
\usepackage{booktabs}
\usepackage{hyperref}

\usepackage[table]{xcolor}
\crefname{figure}{fig}{figures}
\Crefname{figure}{Fig}{Figures}
\begin{document}

\title{Real-World Asset Integration in Next-Generation Communication Networks: Fundamental, Framework, and Case Study}

\author{Tingxuan Su, Haoxiang Luo, Ruichen Zhang, Yinqiu Liu, \\Gang Sun,~\IEEEmembership{Senior Member,~IEEE}, and Hongfang Yu,~\IEEEmembership{Senior Member,~IEEE}
\thanks{T. Su, H. Luo, G. Sun (Corresponding Author), and H. Yu are with the University of Electronic Science and Technology of China, Chengdu 611731, China (e-mail: sutingxuan@std.uestc.edu.cn, lhx991115@163.com; \{gangsun, yuhf\}@uestc.edu.cn). R. Zhang and Y. Liu are with the Nanyang Technological University, Singapore 639798 (e-mail: ruichen.zhang@ntu.edu.sg; yinqiu001@e.ntu.edu.sg; dniyato@ntu.edu.sg). Tingxuan Su and Haoxiang contributed equally to this work.}}



\maketitle

\begin{abstract}
Next-generation communication networks are characterized by integrated ultra-high reliability, ultra-low latency, massive connectivity, and ubiquitous coverage. However, this paradigm faces significant structural challenges of liquidity and security. Liquidity issues arise from prohibitive upfront costs of network resources, which strain the limited capital and financial flexibility. This also limits the deployment of the resource- and investment-intensive security solutions, bringing security issues. Security vulnerabilities arise from the decentralized architecture as well, particularly threats posed by Byzantine nodes. To address these dual challenges, we propose a novel framework utilizing Real-World Asset (RWA) tokenization for tokenizing network resources. RWA tokenization uses blockchain to convert ownership rights of real-world assets into digital tokens that can be programmed, divided, and traded. We then analyze the criteria for identifying suitable assets. Through a case study on dynamic spectrum allocation, we demonstrate the superior performance of this RWA approach. Particularly under conditions of resource scarcity, it can exhibit strong resilience against collusion and default attacks. Finally, we delineate fruitful avenues for future research in this nascent field.
\end{abstract}


\section{Introduction} \label{sec-I}

\IEEEPARstart {N}{ext-generation}
communication networks represent a future network paradigm that integrates ultra-high reliability, ultra-low latency, massive connectivity, and ubiquitous coverage~\cite{next}. This vision supports a wide range of advanced applications, from immersive metaverse environments to the industrial Internet of Things (IoT)~\cite{luo2025weighted}. Compared to conventional networks, next-generation networks offer revolutionary advantages in terms of data rates, connection density, and coverage breadth. Therefore, the next-generation networks can provide an unprecedented technological foundation for true human-machine-intelligent interconnectivity~\cite{Luonext}. However, their realization is impeded by several structural contradictions in their design and deployment, which can be broadly categorized into challenges of \emph{liquidity} and \emph{security}.

\subsubsection{Liquidity Challenges} Some network resources, such as high-performance servers, enterprise-grade cloud computing services, and specialized big data platforms, come with a very high acquisition cost \cite{brinton2025key}. For ordinary individual users or small startup teams, purchasing these resources directly entails significant upfront capital and ongoing operational expenses. It will create a prohibitive economic barrier. This substantial capital expenditure not only depletes their limited financial reserves but also severely restricts their liquidity and financial flexibility. Thus, they are likely to abandon purchasing them, leading to possible resources being idle. 

\subsubsection{Security Challenges} As mentioned above, the high acquisition cost can also limit the deployment of resource- and investment-intensive security solutions, which remain unaffordable for smaller players already struggling with infrastructure costs. Consequently, critical protections can be forgone, leaving these users vulnerable and creating systemic security gaps across the network. Besides this, the architectural shift towards decentralized and collaborative networks introduces complex security threats, most notably those posed by Byzantine nodes. The vast array of heterogeneous network nodes will belong to diverse and potentially adversarial trust domains. Some of these nodes may exhibit malicious or unpredictable behaviors, aiming to disrupt consensus, spoof data, or degrade service \cite{zhu2026moe}. Without robust protective mechanisms, the overall security, collaborative efficiency, and systemic trustworthiness of the network ecosystem can be severely compromised. It results in valuable network resources and financial loss.

To systematically address these dual challenges, Real-World Asset (RWA) tokenization technology emerges as a promising solution. RWA tokenization is an innovative framework that leverages blockchain to transform rights to physical world assets into programmable, divisible, and tradable digital instruments~\cite{RWAExploration}. Specifically, this paper focuses on network resources. Its core advantages, which include transactional transparency, cryptographic security, and the creation of high-liquidity markets, align directly with the needs of the next-generation networks.
On the one hand, RWA tokenization offers a potential solution to the liquidity challenge. It facilitates the fractional ownership and trading of network infrastructure and resources. This mechanism fosters a decentralized global capital market, thereby lowering barriers to entry for distributed investors. On the other hand, it can reduce the significant upfront funding pressures traditionally borne by centralized operators. Concurrently, RWA contributes to addressing security challenges. It leverages architectural properties of blockchain to enhance information traceability. Thus, it can provide superior data transparency and security compared to conventional systems~\cite{food}. When integrated with verifiable digital identities and resource credentials, this combination establishes a cryptographically secure and fault-tolerant trust layer for the network ecosystem.


Considering the pivotal role of RWA tokenization in enabling secure and highly liquid communication networks, this paper explores its potential applications in next-generation communication systems. First, we trace the evolution from Non-Fungible Tokens (NFTs) to RWAs. Next, we propose an architectural framework for network resources RWA and a systematic categorization of tokenizable assets. Finally, we demonstrate the practical effectiveness of this approach and outline emerging directions for future research. The main contributions of this paper are as follows:

\begin{itemize}
    \item We propose an architectural framework for RWA tailored to network resources. By enabling assets to be on-chain and tokenizing them, this approach significantly enhances the liquidity and transactional security of network resources. Furthermore, we introduce two distinct mechanisms, leasing and purchasing, to accommodate diverse needs in specific situations.

    \item We investigate and summarize three principles applicable to assets suitable for RWA tokenization, and identify and analyze which network resources are appropriate for RWA applications.

    \item We explore and demonstrate the effectiveness of the RWA-based approach, highlighting the advantages of this solution with the spectrum resources as a case study.
\end{itemize}

\section{The Evolution from Digital Collectibles to Programmable Assets: From NFTs to RWAs}\label{sec-II}
\begin{figure*}[t]
    \centering
    \includegraphics[width=0.95\textwidth]{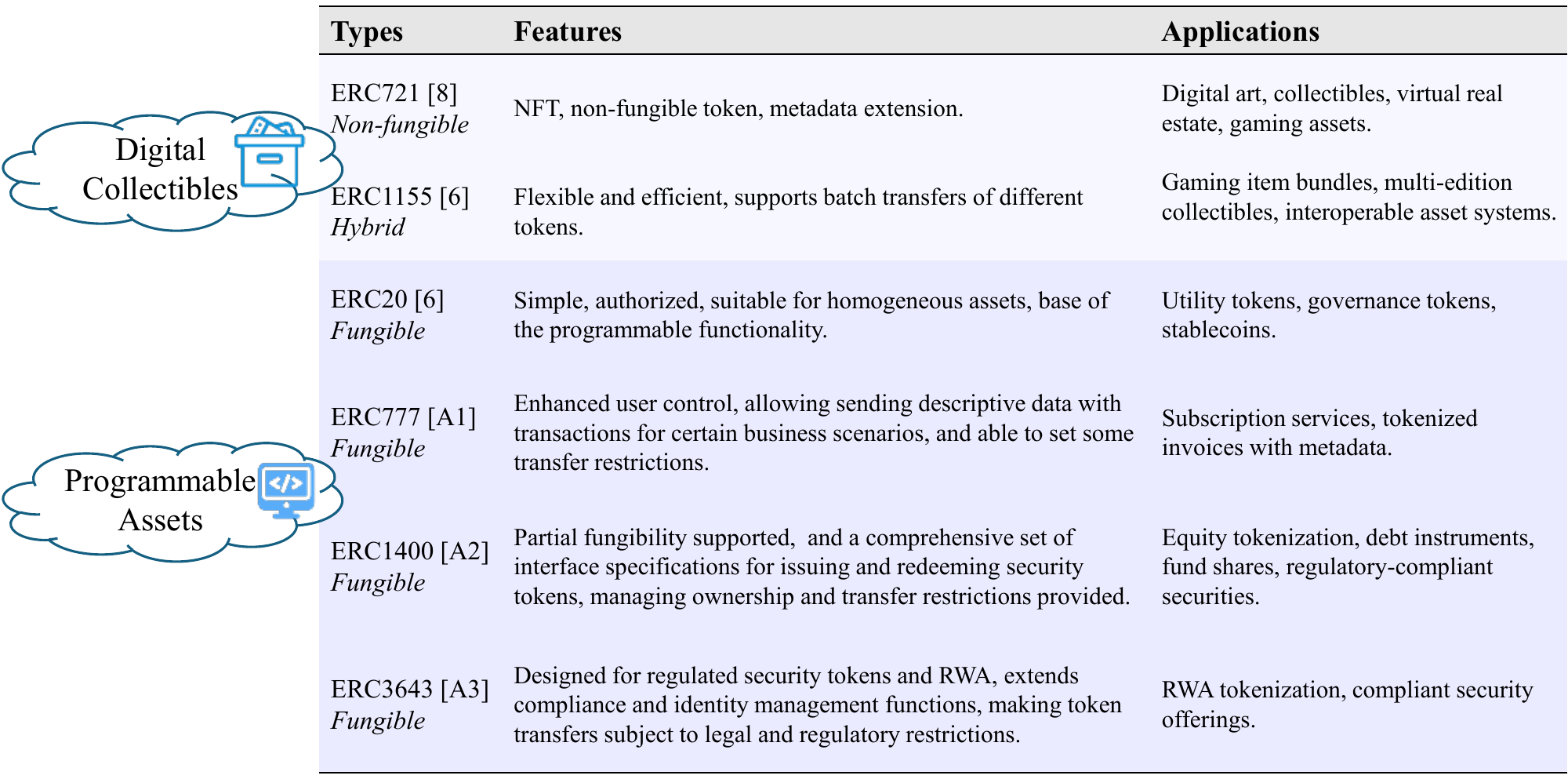}
    \caption{Different types of token standards, with the corresponding substitute features and applications, from NFTs to RWA. Related reference of [A1-A3] can be found in https://eips.ethereum.org/EIPS/eip-777, https://eips.ethereum.org/EIPS/eip-1450, https://eips.ethereum.org/EIPS/eip-3643 respectively.}
    \label{fig:standard}
     \vspace{-0.5cm}
\end{figure*}

NFTs are a class of digital assets built on blockchain, characterized by uniqueness and indivisibility. It enables them to signify exclusive ownership of both digital and physical assets. Originating from standards such as ERC-721~\cite{NFT} and ERC-1155~\cite{RWAExploration} on the Ethereum blockchain, NFTs utilize smart contracts to record ownership and track transactions. Each NFT possesses a distinct identifier and associated metadata, differentiating it from other assets of the same category. 

However, non-programmable NFTs suffer from limitations such as functional staticity, weak interactivity, and a lack of value correlation. Therefore, it cannot dynamically record asset states, interact with external systems, or participate in complex financial operations. 
To address these shortcomings,~\cite{NFToptimization} proposes a method of binding NFTs to accounts. Each NFT is assigned a unique, programmable smart contract address. It endows NFTs with the ability to execute transactions autonomously, hold on-chain assets, and interact with Decentralized Applications (DApps)~\cite{NFToptimization}. 
Nevertheless, such enhancements still face challenges related to technical compatibility, regulatory compliance, and cross-chain interoperability. To realize digital assets that are programmable, interactive, and composable, standards for fungible tokens and hybrid token standards have gradually gained prominence. 

As Fig. \ref{fig:standard} exhibits, the evolutionary progression of token standards culminates in specialized frameworks designed for regulated digital assets.
It begins with ERC-20, a foundational and widely adopted standard for simple, authorized, and fungible tokens representing homogeneous assets~\cite{RWAExploration}. It was followed by ERC-721, which introduced the concept of NFTs with unique metadata extensions, enabling the tokenization of distinct digital items~\cite{NFT}. To enhance user control over fungible tokens, ERC-777 was developed. It allows descriptive data to be attached to transactions and enables configurable transfer restrictions.
A significant leap in flexibility and efficiency arrived with ERC-1155~\cite{RWAExploration}. It supports batch transfers of different token types and combines both fungible and non-fungible assets within a single contract. 
ERC-3643 emerges as the first Ethereum-based compliance token standard purpose-built for RWAs. It extends ERC-20 by integrating robust identity and compliance features.

Introducing RWA on-chain enables the mapping and anchoring of physical assets to the blockchain, thereby infusing digital resources with tangible value, liquidity, and tamper-proofness. In \cite{car}, Notheisen et al. focused on automobile trading scenarios and constructed a layered system based on Ethereum. 
Another approach targeting agricultural assets, such as beef cattle~\cite{NFToptimization}, based on a Directed Acyclic Graph (DAG) blockchain forms an end-to-end system spanning from asset digitization to high-performance trading. 
It provides a viable pathway for the on-chain circulation of low-liquidity physical assets. 
Furthermore, \cite{luo2026real} tokenizes the idle computing resources of drones, electric vertical takeoff and landing vehicles (eVTOLs), and other low-altitude devices as RWA, managing them through blockchain technology. However, the first two remain high-level conceptualizations. Including scheme presented in\cite{luo2026real}, they all fail to account for the fact that network resources are typically not traded directly but leased during idle periods. 

\section{The Architecture for Asset Tokenization} \label{sec-III}
\begin{figure*}[tbp]
    \centering
    \includegraphics[width=0.9\textwidth]{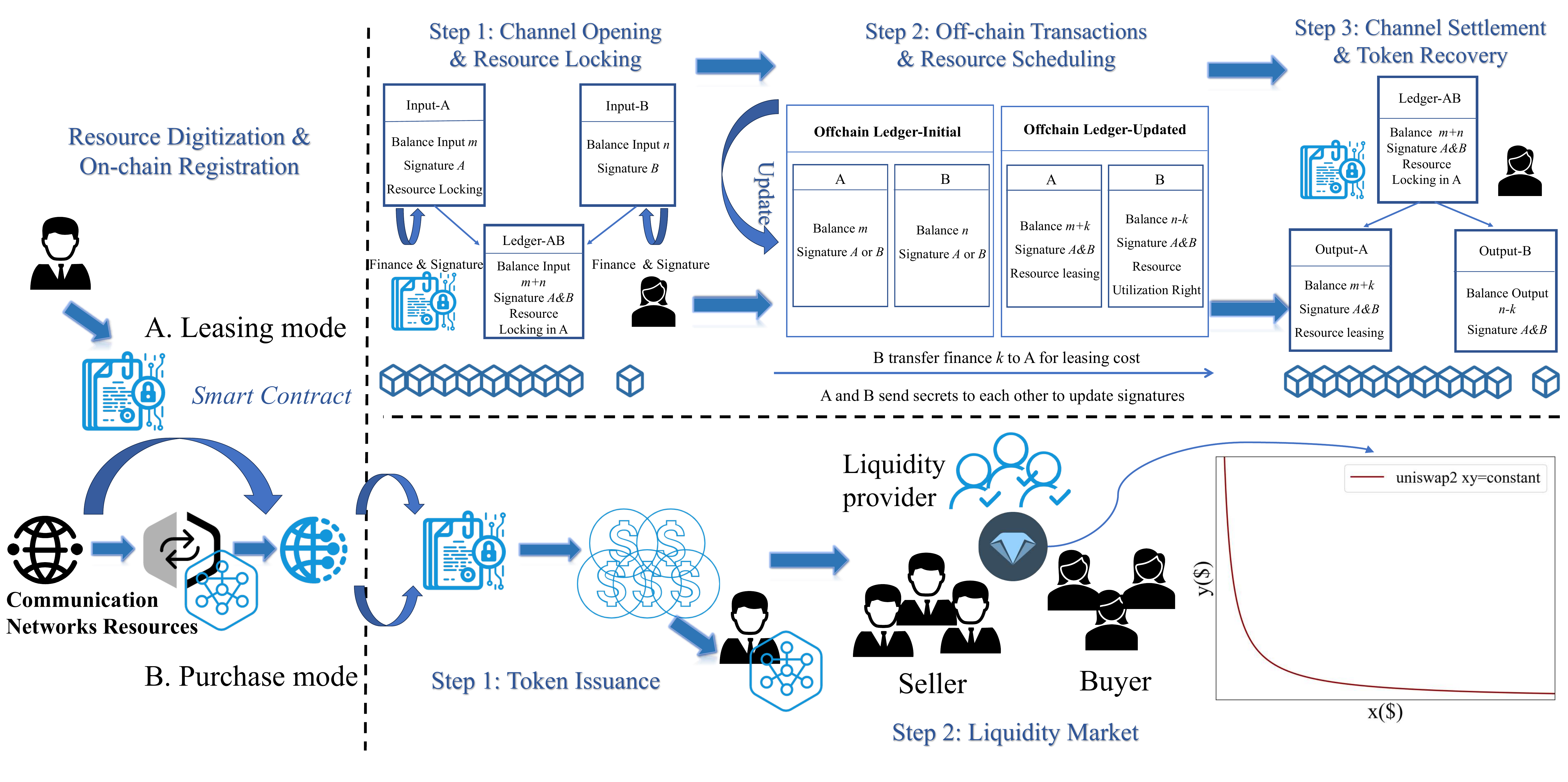}
    \caption{The proposed RWA framework for the next-generation communication networks resources. This framework can enhance the network resource liquidity with high security guaranteed. It also supports two modes with on-demand leasing and real-time trading to accommodate diverse demands. The purchase mode establishes permanent ownership and a liquid market for resource tokens, while the leasing mode enables flexible, low-cost access to resource usage. Together, they form a synergistic economic system where long-term asset value and short-term utility demands reinforce each other, driving an efficient and open decentralized resource ecosystem.}
    \label{fig:overall}
     \vspace{-0.5cm}
\end{figure*}

In this section, we propose a network resource management architecture to assess the applicability RWA, shown in Fig. \ref{fig:overall}. 
\\

Physical network resources are first collected by an owner to extract metadata and performance characteristics, as well as the value evaluated by the owner. They are then encoded into structured digital descriptions. Furthermore, we use smart contracts to create digital twin entities corresponding to the resources on the blockchain, recording their unique identifiers, static attributes, and real-time status~\cite{NFToptimization}. Simultaneously, on-chain accounts are generated for resource owners, completing the initial anchoring and registration of resource property and usage rights.
After the resources are digitized and registered on the blockchain, we will divide this framework into leasing and purchasing to meet different needs.

\begin{itemize}
    \item \textbf{Leasing Mode:} Based on state channels, achieving temporary and flexible transfer of resource usage rights.
    \item \textbf{Purchase Mode:} Achieving long-term transfer of resource usage rights through decentralized markets.
\end{itemize}

\subsection{Leasing Mode}
\textbf{\textit{Step 1: State Channel Opening and Resource Locking:}}
Resource owners and lessees can initiate a resource leasing intent via on-chain transactions. After verifying the identities and token ownership of both parties, the smart contract automatically deploys an off-chain state channel. It temporarily binds the usage rights of the corresponding resource to the tokens. The state channel records the lease duration, service level agreement, and payment conditions, while locking the resource tokens for circulation within the channel. 

\textbf{\textit{Step 2: Off-chain Transactions and Resource Scheduling:}}
Within the state channel, lessees can achieve fine-grained transactions of resource usage rights through token transfers or payment sub-channels. The real-time status of resource scheduling is updated via consensus among channel participants through signed messages, without requiring on-chain submission for each update. 
The final state is submitted to the on-chain contract only in case of disputes or settlement.

\textbf{\textit{Step 3: Channel Settlement and Token Recovery:}}
Upon lease expiration or voluntary termination by the lessee, the state channel enters the settlement phase. The smart contract verifies the final state of the channel and automatically executes token recovery and rent distribution. Ownership of the resource tokens remains with the original owner, while the revenue generated during the lease period is converted into stablecoins or native tokens and distributed to the owner in accordance with the agreed-upon ratio. 

\subsection{Purchase Mode}

\textbf{\textit{Step 1: Token Issuance:}}
Smart contracts divide long-term resource usage rights into homogeneous tokens and bind them to the owner's account. 

\textbf{\textit{Step 2: Liquidity Market:}}
Resource tokens can be traded on decentralized exchanges through AMM mechanisms, forming a liquidity market. Token prices are dynamically adjusted based on resource valuation and market supply and demand.
Buyers can directly acquire ownership of resource tokens through the market and freely decide whether to use them for subsequent leasing, holding, or resale.

\subsection{Benefits from the RWA framework}
\begin{figure}[t]
    \centering
    \includegraphics[width=0.49\textwidth]{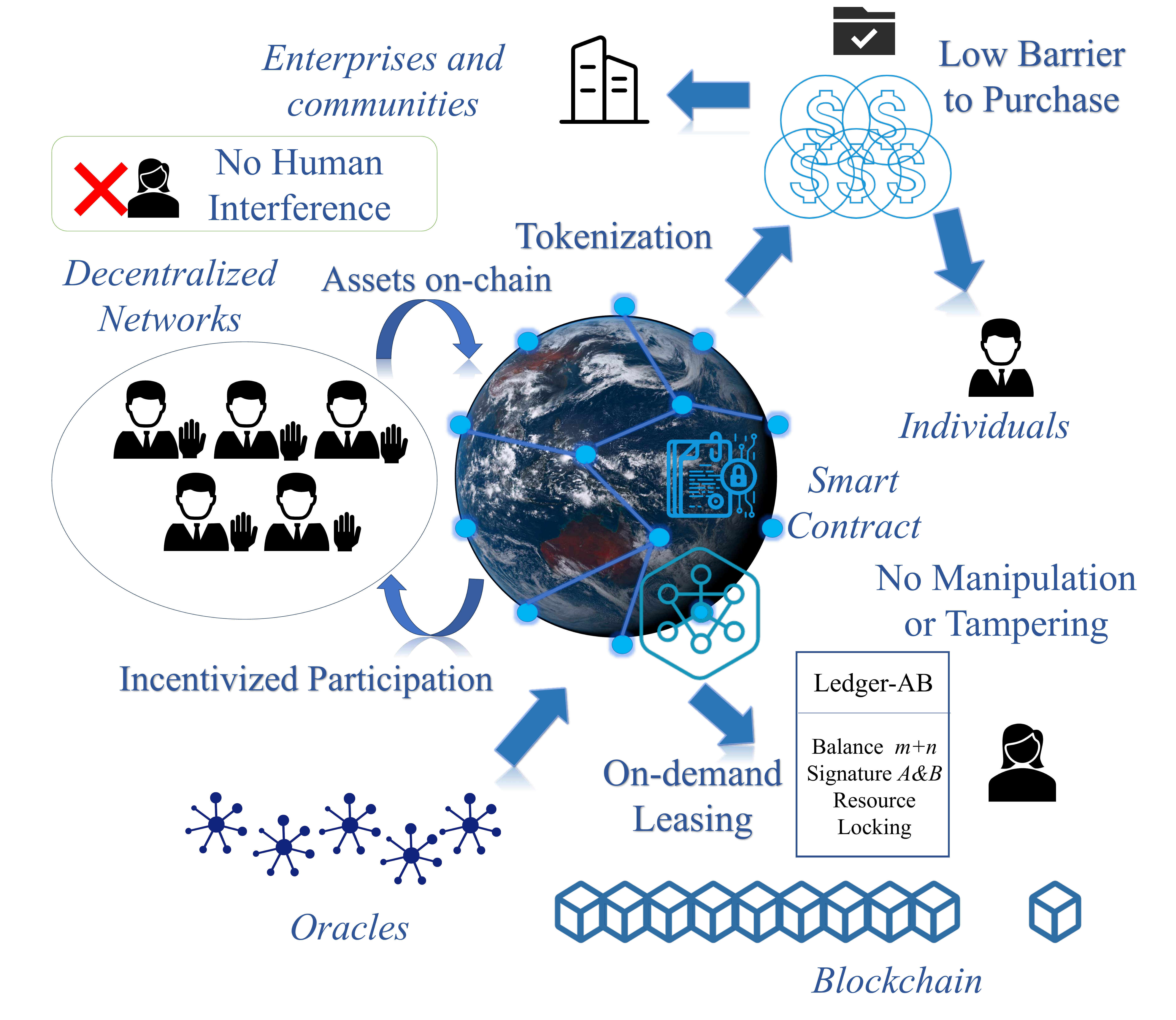}
    \caption{Network resources benefits from the proposed RWA framework. It can help enhance market liquidity and lower barriers to entry, strengthen security and trustless operations, and incentivize mass participation and decentralized
    network growth.}
    \label{fig:benefits}
     \vspace{-0.5cm}
\end{figure}

As Fig. \ref{fig:benefits} exhibits, the tokenization of the aforementioned resources as RWAs can yield the following significant benefits:
\begin{itemize}
\item \textbf{Enhanced Market Liquidity and Lowered Barriers to Entry:}
Traditional resource allocation is often inefficient and exclusive. 
RWA tokenization transforms resources into fractional, tradable shares, realizing on-demand leasing and real-time trading. This creates a liquid marketplace. The idle resources, such as spare compute power or underutilized bandwidth, can be dynamically priced and allocated based on real-time supply and demand. 

\item \textbf{Strengthened Security and Trustless Operations:}
Representing network resources as tokenized RWAs allows operational rules to be encoded directly into self-executing smart contracts, eliminating reliance on intermediaries.
Resource states are securely and transparently synchronized on-chain via oracles, enabling continuous, verifiable monitoring and automated compliance. The immutability of blockchain records ensures that all transactions and usage histories are auditable and tamper-proof. 

\item \textbf{Incentivized Mass Participation and Decentralized Network Growth:}
RWA tokenization democratizes network ownership and contribution. By fractionalizing large-scale infrastructure
it helps small enterprises, communities, and individuals to own stakes and earn yields from previously inaccessible assets. Simultaneously, anyone with idle resources like home bandwidth or spare computing cycles can tokenize and contribute them to the network, receiving economic incentives in return. 
\end{itemize}

The leasing and purchase mode 
complement each other and cater to different market demands and participant types. The purchase mode establishes the long-term value benchmark and liquidity depth for resource assets through AMM mechanisms. It can attract investors and long-term builders, thereby laying a solid foundation of property rights and capital for the network. Meanwhile, the leasing mode, leveraging state channels, achieves the flexible, efficient, and low-cost circulation of resource usage rights. 
Operating in synergy, they jointly drive the transformation of network resources from rigid asset possession to efficient and open value circulation.

\begin{table*}[htbp]
\centering
\caption{Analysis of Characteristics for Tokenizing Digital Resources as RWAs}
\label{tab:rwa_analysis}

\begin{tblr}{
  width = \linewidth,
  colspec = {
    Q[l,bg=gray!10,m] 
    Q[l,m]            
    Q[l,m]             
    Q[l,m]             
  },
  column{1} = {wd=2.6cm},  
  column{2} = {wd=2.9cm},    
  column{3} = {wd=5.2cm},    
  column{4} = {wd=5.55cm},  
  row{1} = {bg=gray!20, font=\bfseries}, 
  row{2} = {bg=blue!3},    
  row{3} = {bg=blue!8},    
  row{4} = {bg=blue!3},    
  row{5} = {bg=blue!8},    
  row{6} = {bg=blue!3},    
  row{7} = {bg=blue!8}, 
  row{8} = {bg=blue!3},        
  hline{1} = {2pt},        
  hline{2} = {1pt},     
  hline{Z} = {2pt},      
  vline = {},          
  cell{2-9}{1} = {bg=gray!10, font=\bfseries}, 
}
Resource Category & Attribute Classification & Core Advantages for Tokenization & Primary Challenge \\
Domain Name & Intangible Asset (Intellectual Property) & \textbf{Inherently unique, akin to NFTs}. Mature market with clear ownership, facilitating easy on-chain mapping and circulation. & Subject to national policies and regulations, and there is currently a lack of a clear regulatory framework for its securitization as a standardized tradable financial asset. \\
Internet Protocol (IP) Address Block & Intangible Asset (Usufruct) & \textbf{A naturally scarce digital resource}. Can generate stable cash flow through leasing, suitable for yield-bearing tokens. & Legally recognized as a right of use in most jurisdictions instead of ownership, creating complex compliance and ownership-structuring issues for tokenization. \\
Network Bandwidth & Infrastructure (Telecommunications) & \textbf{Enables fractional ownership of high-value dedicated lines}, lowering investment barriers. Supported by stable enterprise demand. & Asset value is heavily reliant on physical infrastructure and carrier performance, making off-chain verification difficult. \\
Node Resources & Intangible Asset (Node Access/Permission) & \textbf{Access or control permissions for core network nodes hold potential strategic value.} Can grant the power to deliver superior service to insiders and secure greater commercial profits. & High risk of violating telecommunications regulations and net neutrality principles, posing significant compliance hurdles. \\
Computational Power & Infrastructure (Computing) & \textbf{Highly standardized and divisible}. Driven by clear and growing market demand. Rights distribution can be automated via smart contracts. & Value is significantly impacted by hardware depreciation, energy costs, and maintenance quality, requiring trusted performance verification. \\
Radio Spectrum & Infrastructure (Telecommunications) & \textbf{Physical scarcity guarantees long-term value}. Long-term licenses are suitable for structuring as yield-generating assets. & Vastly differing regulatory policies across countries create fundamental barriers for cross-border tokenization and trading. \\
Cloud Storage Resources & Infrastructure (Computing) & \textbf{Standardized, granular, and high-demand}. Enables micro-monetization of idle capacity. & Performance hard to verify; value tied to hardware specs, depreciation, and data privacy. \\
\end{tblr}
\end{table*}

\section{Identifying Tokenizable Assets in Communication Networks} \label{sec-IV}

Employing RWA technology can achieve the tokenization of nearly all asset types~\cite{NFToptimization}. However, research indicates that not all resources are equally suitable for this process. Ideal candidate assets for tokenization should satisfy three fundamental criteria:

\begin{itemize}
\item \textbf{Intrinsic and Stable Value:} The asset should primarily possess long-term stable value characteristics. Its economic worth must be derived from physical properties, substantive utility, or contractual cash flows, rather than from speculative future income projections~\cite{NFToptimization}. 

\item \textbf{Strategic Economic Scarcity:} The asset should hold significant strategic importance or perform critical functions within its economic sector. Its value proposition must be underpinned by verifiable scarcity, non-fungibility, or unique functional attributes that create sustainable economic moats~\cite{security}. Such characteristics ensure the asset maintains relevance and demand within markets.

\item \textbf{Legal Clarity and Fungibility:} The asset should operate within an unambiguous legal framework with clearly defined ownership rights or at least usage rights.
It should be amenable to standardization or divisible into structured, legally recognized units. A definitive separation between ownership rights and usage rights is essential to achieve high transferability. 
\end{itemize}

Based on the summarized criteria, seven types of network resources can be recognized as shown in TABLE \ref{tab:rwa_analysis}. As it shows, while many resources exhibit strong alignment with the criteria of intrinsic value and strategic economic importance, they frequently face significant hurdles regarding the third criterion of legal clarity and fungibility. Challenges such as unclear regulatory frameworks, complex ownership/rights structures, and jurisdiction-specific policies are predominant. This indicates that legal and compliance barriers, rather than technical or economic feasibility, often constitute the primary constraint for their successful tokenization as RWAs \cite{RWAExploration}.

\section{Case Study} \label{sec-VI}
\subsection{Case Scenario}
We use the radio spectrum as a typical example for this study. In the traditional spectrum allocation model, radio spectrum, a finite and strategically vital physical resource, is typically licensed by regulatory bodies to large telecommunications operators through lengthy and opaque administrative or auction processes~\cite{spectrum}. This creates high barriers to entry, leading to underutilization of licensed frequency bands and resulting in a rigid, illiquid market. In this market, spectrum assets are isolated and difficult to dynamically share or trade.

In our case study, we envision a scenario provided by RWA tokenization. Discrete licensed spectrum blocks of a 10 MHz bandwidth in the 3.5 GHz band are tokenized as RWAs on a blockchain. Each spectrum license with one hundred 0.1 MHz spectrum slices is decomposed into standardized, divisible ownership tokens compliant with the ERC-3643\footnote{https://eips.ethereum.org/EIPS/eip-3643} security token standard. It also represents a share of the usage rights. 
We use the smart contract to dynamically generate usage right tokens, which correspond to authorized access to a specific slice for the defined period, with a minimum granularity of one hour. Using the AMM mechanism~\cite{NFToptimization}, this scheme creates a liquid digital asset that can be freely but compliantly traded. It is transforming the spectrum from a static administrative asset into a dynamic, financialized commodity.


\subsection{Experimental Setup}
The performance of the proposed RWA scheme was evaluated through agent-based simulations. The simulations are implemented in Python using the Mesa framework within the VS Code IDE and conducted on a computer equipped with an Intel(R) Xeon(R) Gold 5220R CPU @ 2.20 GHz. Interactions with the blockchain were simulated using a 5-second block timeout and a block size of 10 transactions.
Three typical schemes in edge computing resource and transaction scheduling are selected for comparison:
\begin{itemize}
    \item \textbf{Matching and Pricing Resource Allocation (MPRA)~\cite{multiparticipant}:} 
    For long-term stable markets, it uses ranking matching and binary search pricing to prioritize large bid-ask spread parties.
    \item \textbf{Truthful Resource Allocation (TRA)~\cite{multiparticipant}:} For short-term dynamic markets, it follows the lowest-price seller first, equal allocation, and minimum viable bid principles to encourage truthful reporting and improve efficiency
    \item \textbf{Competitive Padding Auction (CPA)~\cite{consortium}:} A blockchain-integrated competitive auction mechanism. It achieves decentralized execution via consortium chain and smart contracts, realizes asymptotic economic efficiency, 
    and supports peer-to-peer energy trading.
    
\end{itemize}


To give a fair comparison focused on the core market mechanisms and to highlight the advantages of our proposed design, we exclude the leasing mode, which is absent in these traditional schemes, and focus solely on purchase and sale transactions.
We then test both liquidity and security, where resource liquidity is defined as resource utilization, which means that sold resources divided by initial total resources. Specifically, we test resource utilization as the number of buyers changes while keeping the number of sellers fixed at 100. This means we are essentially examining how different schemes perform as resources become increasingly scarce with a growing buyer population. For security tests, we set the proportion of Byzantine nodes from 0 to 30\% and with taking into account the requirement for 30\% node tolerance in BFT protocols~\cite{luo2025weighted}. Under this security baseline, we tested the system's resilience under three types of market attacks. As shown below:

\begin{figure}[t]
    \centering
    \includegraphics[width=0.46\textwidth]{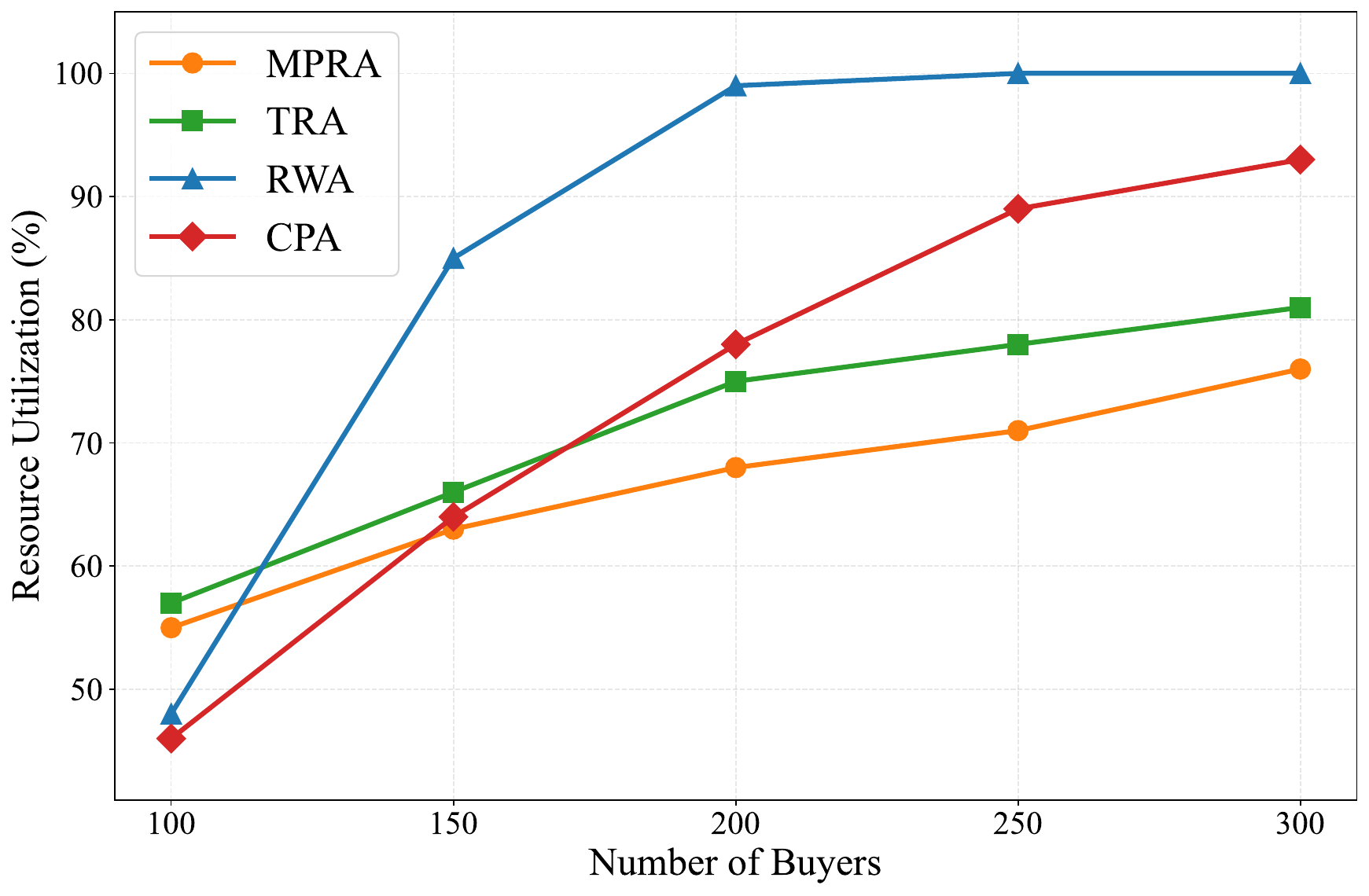}
    \caption{Utilization comparison among RWA, MPRA, TRA, CPA. Utilization varies with the number of buyers ranging from 100 to 300 with the number of sellers fixed at 100.}
    \label{fig:buyers}
     \vspace{-0.5cm}
\end{figure}

\begin{figure*}[t]
    \centering
    \includegraphics[width=0.99\textwidth]{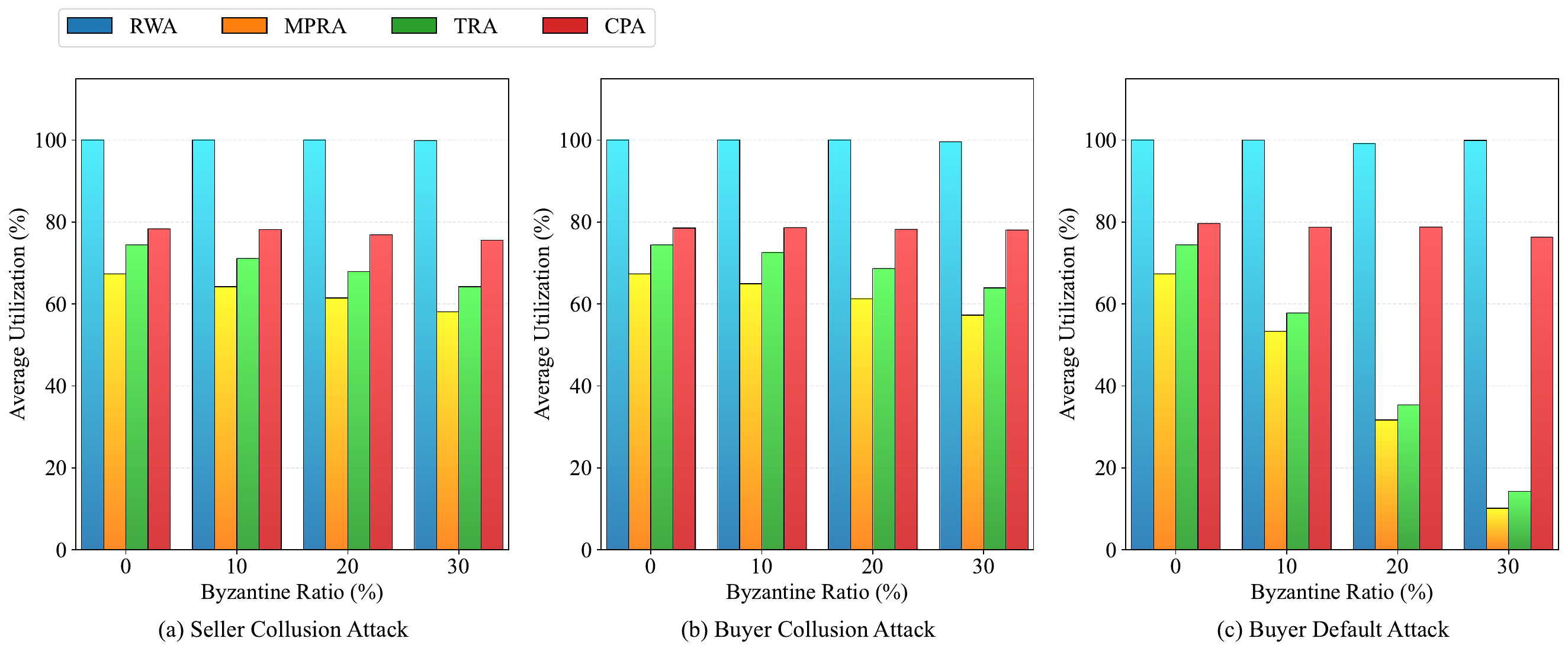}
    \caption{Utilization comparison among RWA, MPRA, TRA, and CPA schemes under varying Byzantine node ratios and attack scenarios. The experiments are conducted with a fixed configuration of 200 buyers and 100 sellers. (a) Performance under buyer collusion attacks, where malicious buyers coordinate to submit artificially low bids to depress market prices. (b) Performance under seller collusion attacks, where malicious sellers coordinate to inflate asking prices and restrict supply to drive up prices. (c) Performance under default attacks, where malicious buyers probabilistically refuse payment after obtaining resources, thereby disrupting transaction integrity.}
    \label{fig:byzantine_rare}
     \vspace{-0.5cm}
\end{figure*}

\begin{itemize}
    \item \textbf{Buyer Collusion:} It is a systematic market manipulation by multiple coordinated malicious nodes. It involves multiple malicious buyers collectively depressing their bids to distort the demand curve, aiming to drive market prices down abnormally.
    \item \textbf{Seller Collusion}: It involves multiple malicious sellers jointly inflating asking prices and controlling supply to artificially create scarcity and drive prices up. 
    \item \textbf{Default attack:} It is a post-transaction fraudulent behavior where malicious buyers probabilistically refuse payment after obtaining resources. Its core harm is the direct disruption of the value exchange loop, reducing both statistical and economic resource utilization.
\end{itemize}   



\subsection{Experimental Performance}
\subsubsection{Liquidity Discussion}

As shown in Fig. \ref{fig:buyers}, the resource utilization of all schemes increases with the number of buyers. This occurs because when sellers outnumber buyers, resources are abundant and remain undersubscribed. As buyers gradually increase, resource utilization rises accordingly. By leveraging the RWA technique to tokenize spectrum resources into finer-grained units, the proposed scheme achieves higher spectrum allocation granularity. Consequently, for the same number of buyers and sellers, it delivers faster performance improvements and superior overall efficiency. When the number of nodes exceeds 100, the RWA scheme significantly outperforms other schemes. When the number of buyers reaches 200, the demand exactly matches the supply, leading to 100\% spectrum utilization. However, it can be observed that at the point of 100 buyers, the RWA-based schemes underperform compared to MPRA and TRA. This can be attributed to the relatively abundant resources under such conditions. In the RWA simulations, since spectrum resources are considered sold only when all fragmented units are purchased as a whole, some residual unsold fragments may remain pending, commonly referred to as leftover orders. 
Therefore, we can conclude that when spectrum resources are relatively scarce, the proposed RWA scheme performs significantly better than the other three benchmarks. This outcome aligns with the initial objective. It addresses the gap between the urgent demand for efficient dynamic resource allocation and the limitations of current static mechanisms by enabling finer-grained and more liquid resource allocation scenarios.





\subsubsection{Security Analysis}
As illustrated in Fig. \ref{fig:byzantine_rare}, when the number of sellers is 100 and that of buyers is 100, the utilization rate of RWA remains close to 100\% without significant decline under the three types of attacks. In contrast, the other three comparative methods exhibit noticeable decreases. Collusion-based attacks aim to influence the market through bid manipulation. RWA, however, determines prices directly via an AMM formula, with liquidity providers continuously smoothing the price curve, rendering it largely insensitive to bid-based attacks. In contrast, collusion among buyers and sellers may lead to the following effects respectively: for sellers, the minimum feasible bid may be depressed, causing some sellers to exit the market because prices fall below costs; for buyers, the minimum feasible asking price may be inflated, resulting in buyers being unable to match with suitable orders. Both situations will lead to an overall decrease in resource utilization. CPA, by virtually incorporating quotation nodes into its optimization process, also demonstrates a degree of resilience against such attacks. In scenarios involving buyer collusion, for CPA, because funds unused in the current time step are carried over to the next, no significant changes are observed, as reflected in Fig. \ref{fig:byzantine_rare}. In the case of default attacks, MPRA and TRA show marked declines in resource utilization, whereas RWA and CPA, strengthened by blockchain-related mechanisms, exhibit robust security and effectively resist default attacks.


\section{Future Research Directions} \label{sec-VII}
\emph{\textbf{1. Adaptive Smart Contracts:}}
The RWA solution can effectively eliminate third-party intervention, thereby reducing costs. However, its upfront deployment expenses remain relatively high, particularly in areas such as the development, auditing, and deployment of smart contracts. 
Therefore, integrating artificial intelligence technologies to develop adaptive smart contracts represents a highly promising direction. Such contracts could autonomously adjust their parameters or logic in response to dynamic network conditions, market fluctuations, or predefined performance thresholds, thereby reducing the frequency of manual interventions and costly redeployments.

\emph{\textbf{2. On-chain Costs:}}
Consensus and storage mechanisms of blockchain systems, significant latency and on-chain overhead are inevitably introduced. The costs associated with subsequent on-chain operations must be carefully evaluated during system design and economic modeling. The efficiency and cost issues during the on-chain execution phase likewise require systematic consideration. 

\emph{\textbf{3. Privacy Balances:}}
In the implementation of network resource RWAs, privacy considerations present a fundamental dilemma. 
Purchasing patterns of computing power may reveal research and development directions, spectrum transaction data can infer group activity patterns, and storage access logs may conceal business secrets. 
The future solutions will need to be achieved through privacy-enhancing technologies, such as zero-trust architecture, to achieve controllable transparency.

\section{Conclusion} \label{sec-VIII}
This paper presents the challenges, foundational solutions, and a concrete use case for integrating RWA tokenization into next-generation networks. We have highlighted that conventional network architectures are fundamentally impeded by structural contradictions in liquidity and security. The proposed blockchain-based RWA framework directly addresses these issues by enabling fractional, programmable, and verifiable tokenization of network resources, thus creating a liquid marketplace for infrastructure and establishing a Byzantine fault-tolerant trust layer. We have presented a case study on dynamic spectrum allocation to demonstrate the superior performance of the proposed RWA-based scheme. Particularly under conditions of resource scarcity, it significantly outperforms other mechanisms in utilization while maintaining robust resilience against collusion and default attacks. Finally, we have discussed future research directions.

\bibliographystyle{IEEEtran}
\bibliography{IEEEabrv,mylib}



\vspace{3em}

\vfill

\end{document}